\documentclass[pra,aps,nopacs,onecolumn,twoside,superscriptaddress]{revtex4}

\usepackage{amsmath,amsfonts,amssymb,caption,color,epsfig,graphics,graphicx,hyperref,latexsym,mathrsfs,revsymb,theorem,url,verbatim,epstopdf,mathtools,enumerate,tikz}
\usetikzlibrary{matrix}
\usetikzlibrary{decorations.pathreplacing}
\hypersetup{colorlinks,linkcolor={blue},citecolor={blue},urlcolor={red}}

\newtheorem{definition}{Definition}
\newtheorem{proposition}[definition]{Proposition}
\newtheorem{lemma}[definition]{Lemma}

\newtheorem{theorem}[definition]{Theorem}
\newtheorem{corollary}[definition]{Corollary}
\newtheorem{conjecture}[definition]{Conjecture}

\newtheorem{remark}[definition]{Remark}
\newtheorem{example}[definition]{Example}
\newtheorem{question}[definition]{Question}
\newtheorem{memo}[definition]{Memo}

\def\squareforqed{\hbox{\rlap{$\sqcap$}$\sqcup$}}
\def\qed{\ifmmode\squareforqed\else{\unskip\nobreak\hfil
\penalty50\hskip1em\null\nobreak\hfil\squareforqed
\parfillskip=0pt\finalhyphendemerits=0\endgraf}\fi}
\def\endenv{\ifmmode\;\else{\unskip\nobreak\hfil
\penalty50\hskip1em\null\nobreak\hfil\;
\parfillskip=0pt\finalhyphendemerits=0\endgraf}\fi}
\newenvironment{proof}{\noindent \textbf{{Proof.~} }}{\qed}
\def\Dbar{\leavevmode\lower.6ex\hbox to 0pt
{\hskip-.23ex\accent"16\hss}D}
\makeatletter
\def\url@leostyle{%
  \@ifundefined{selectfont}{\def\UrlFont{\sf}}{\def\UrlFont{\small\ttfamily}}}
\makeatother
\def\bcj{\begin{conjecture}}
\def\ecj{\end{conjecture}}
\def\bcr{\begin{corollary}}
\def\ecr{\end{corollary}}
\def\bd{\begin{definition}}
\def\ed{\end{definition}}
\def\bea{\begin{eqnarray}}
\def\eea{\end{eqnarray}}
\def\beq{\begin{equation}}
\def\eeq{\end{equation}}
\def\bal{\begin{aligned}}
\def\eal{\end{aligned}}
\def\bem{\begin{enumerate}}
\def\eem{\end{enumerate}}
\def\bex{\begin{example}}
\def\eex{\end{example}}
\def\bim{\begin{itemize}}
\def\eim{\end{itemize}}
\def\bl{\begin{lemma}}
\def\el{\end{lemma}}
\def\bma{\begin{bmatrix}}
\def\ema{\end{bmatrix}}
\def\bpf{\begin{proof}}
\def\epf{\end{proof}}
\def\bpp{\begin{proposition}}
\def\epp{\end{proposition}}
\def\bqu{\begin{question}}
\def\equ{\end{question}}
\def\br{\begin{remark}}
\def\er{\end{remark}}
\def\bt{\begin{theorem}}
\def\et{\end{theorem}}
\def\bmm{\begin{memo}}
\def\emm{\end{memo}}

\def\btb{\begin{tabular}}
\def\etb{\end{tabular}}

\newcommand{\nc}{\newcommand}

\nc{\bbA}{\mathbb{A}} \nc{\bbB}{\mathbb{B}} \nc{\bbC}{\mathbb{C}}
 \nc{\bbD}{\mathbb{D}} \nc{\bbE}{\mathbb{E}} \nc{\bbF}{\mathbb{F}}
 \nc{\bbG}{\mathbb{G}} \nc{\bbH}{\mathbb{H}} \nc{\bbI}{\mathbb{I}}
 \nc{\bbJ}{\mathbb{J}} \nc{\bbK}{\mathbb{K}} \nc{\bbL}{\mathbb{L}}
 \nc{\bbM}{\mathbb{M}} \nc{\bbN}{\mathbb{N}} \nc{\bbO}{\mathbb{O}}
 \nc{\bbP}{\mathbb{P}} \nc{\bbQ}{\mathbb{Q}} \nc{\bbR}{\mathbb{R}}
 \nc{\bbS}{\mathbb{S}} \nc{\bbT}{\mathbb{T}} \nc{\bbU}{\mathbb{U}}
 \nc{\bbV}{\mathbb{V}} \nc{\bbW}{\mathbb{W}} \nc{\bbX}{\mathbb{X}}
 \nc{\bbZ}{\mathbb{Z}}

 \nc{\bA}{{\bf A}} \nc{\bB}{{\bf B}} \nc{\bC}{{\bf C}}
 \nc{\bD}{{\bf D}} \nc{\bE}{{\bf E}} \nc{\bF}{{\bf F}}
 \nc{\bG}{{\bf G}} \nc{\bH}{{\bf H}} \nc{\bI}{{\bf I}}
 \nc{\bJ}{{\bf J}} \nc{\bK}{{\bf K}} \nc{\bL}{{\bf L}}
 \nc{\bM}{{\bf M}} \nc{\bN}{{\bf N}} \nc{\bO}{{\bf O}}
 \nc{\bP}{{\bf P}} \nc{\bQ}{{\bf Q}} \nc{\bR}{{\bf R}}
 \nc{\bS}{{\bf S}} \nc{\bT}{{\bf T}} \nc{\bU}{{\bf U}}
 \nc{\bV}{{\bf V}} \nc{\bW}{{\bf W}} \nc{\bX}{{\bf X}}
 \nc{\bZ}{{\bf Z}}

\nc{\cA}{{\cal A}} \nc{\cB}{{\cal B}} \nc{\cC}{{\cal C}}
\nc{\cD}{{\cal D}} \nc{\cE}{{\cal E}} \nc{\cF}{{\cal F}}
\nc{\cG}{{\cal G}} \nc{\cH}{{\cal H}} \nc{\cI}{{\cal I}}
\nc{\cJ}{{\cal J}} \nc{\cK}{{\cal K}} \nc{\cL}{{\cal L}}
\nc{\cM}{{\cal M}} \nc{\cN}{{\cal N}} \nc{\cO}{{\cal O}}
\nc{\cP}{{\cal P}} \nc{\cQ}{{\cal Q}} \nc{\cR}{{\cal R}}
\nc{\cS}{{\cal S}} \nc{\cT}{{\cal T}} \nc{\cU}{{\cal U}}
\nc{\cV}{{\cal V}} \nc{\cW}{{\cal W}} \nc{\cX}{{\cal X}}
\nc{\cZ}{{\cal Z}}

\nc{\hA}{{\hat{A}}} \nc{\hB}{{\hat{B}}} \nc{\hC}{{\hat{C}}}
\nc{\hD}{{\hat{D}}} \nc{\hE}{{\hat{E}}} \nc{\hF}{{\hat{F}}}
\nc{\hG}{{\hat{G}}} \nc{\hH}{{\hat{H}}} \nc{\hI}{{\hat{I}}}
\nc{\hJ}{{\hat{J}}} \nc{\hK}{{\hat{K}}} \nc{\hL}{{\hat{L}}}
\nc{\hM}{{\hat{M}}} \nc{\hN}{{\hat{N}}} \nc{\hO}{{\hat{O}}}
\nc{\hP}{{\hat{P}}} \nc{\hR}{{\hat{R}}} \nc{\hS}{{\hat{S}}}
\nc{\hT}{{\hat{T}}} \nc{\hU}{{\hat{U}}} \nc{\hV}{{\hat{V}}}
\nc{\hW}{{\hat{W}}} \nc{\hX}{{\hat{X}}} \nc{\hZ}{{\hat{Z}}}

\nc{\hn}{{\hat{n}}}

\newcommand{\red}{\textcolor{red}}

\usepackage{hyperref}

\begin{document}

\Large

\title{Reply to Comment on 'Product states and Schmidt rank of mutually unbiased bases in dimension six'}

\author{Mengfan Liang}\email[]{}
\affiliation{LMIB(Beihang University), Ministry of education, and School of Mathematical Sciences, Beihang University, Beijing 100191, China}

\author{Lin Chen}\email[]{linchen@buaa.edu.cn (corresponding author)}
\affiliation{LMIB(Beihang University), Ministry of education, and School of Mathematical Sciences, Beihang University, Beijing 100191, China}

\begin{abstract}
Daniel McNulty $et.al$ \cite{commentMUB} voiced suspicions to the Lemma 11(v) Part 6 in \cite{Chen2017Product} and three theorems
derived in later publications \cite{Liang2019,lma2020,xyc}. For these suspicions, we reprove that any $6\times 6$ complex Hadamard matrix whose number of real elements more than 22 does not belong to a set of four mutually unbiased bases. We show that the number of $2\times 2$ complex Hadamard submatrices of any $H_2$-reducible matrix is not $10,...,16,18$. We also put forward some possible directions for further development.
\end{abstract}

\date{\today}
\maketitle


\section{Introduction}
Generally, 
 $k$ mutually unbiased bases (MUBs) in $d$-dimensional Hilbert space $\bbC^d$ are $k$ orthogonal bases, such that any two vectors from different bases have inner product of modulus $1/\sqrt d$. Only when $k=d+1$, we say the  $k$ MUBs in $\bbC^d$ is complete. As far as we know, the complete MUBs exist in $\mathbb{C}^d$ when $d$ is a prime power \cite{WOOTTERS1989363}. How to find the complete MUBs in $\mathbb{C}^6$ is the first unsolved case and it is also a famous open problem in quantum information. \par
 The suspicions mentioned in \cite{commentMUB} are enumerated as follows.
 \begin{itemize}
     \item (\cite{commentMUB}, Lemma 1) If a set of four MU bases in dimension six exists, none of the Hadamard matrices from the set contains a real $3\times 2$ submatrix.
     \item (\cite{commentMUB}, Theorem 1) If a set of four MU bases in dimension six exists, none of the Hadamard matrices from the set contains more than 22 real entries.
     \item (\cite{commentMUB}, Theorem 2) An $H_2$-reducible matrix in a set of four MU bases contains exactly nine or eighteen $2\times 2$ submatrices proportional to Hadamard matrices.
     \item (\cite{commentMUB}, Theorem 3) An $H_2$-reducible matrix in a set of four MU bases contains exactly nine $2\times 2$ submatrices each proportional to a Hadamard matrix. Thus, members of the families $D_6^{(1)},B_6^{(1)},M_6^{(1)}$ and $X_6^{(2)}$ do not figure in a quadruple of MU bases.
 \end{itemize}
They think that the proof of the lemma in \cite{Chen2017Product} contains a mistake, ultimately invalidating the three theorems derived in later publications \cite{Liang2019,lma2020,xyc}. For these questions, we reprove the Theorem 1. Then we show that Theorems 2 and 3 are equivalent, and they imply Lemma 1. So it suffices to prove Theorem 2 or 3. We also put forward some ideas to investigate Theorem 2 and Lemma 1. Next, we will reply one by one in different sections.

\section{Theorem 1 in \cite{commentMUB}}
For the Theorem 1 mentioned in \cite{commentMUB}, we think that it is still correct, and we provide a new proof. First of all, we introduce the following lemma.
\begin{lemma}
\label{LMF2}
Let $H_6$ be a $6\times 6$ complex Hadamard matrix (CHM) whose number of real elements is $n$. If $n>22$, then $H_6$ is  equivalent to the matrix
\begin{eqnarray}
\label{M23}
\begin{bmatrix}
i & 1 & 1 & 1 & 1 & 1\\
1 & i & 1 & 1 & -1 & -1\\
1 & 1 & i & -1 & 1 & -1\\
1 & 1 & -1 & i & -1 & 1\\
1 & -1 & 1 & -1 & i & 1\\
1 & -1 &-1 & 1 & 1 & i
\end{bmatrix} or \begin{bmatrix}
\omega & \omega & 1 & 1 & 1 & 1\\
\omega & -\omega & -1 & 1 & -1 & 1\\
1 & 1 & \omega & \omega & 1 & 1\\
1 & -1 & -\omega & \omega & -1 & 1\\
1 & 1 & 1 & 1 & \omega &\omega\\
-1 & 1 &1 & -1 & \omega & -\omega
\end{bmatrix},
\end{eqnarray}
where $\omega=e^{\frac{2\pi i}{3}}\ or\ e^{\frac{4\pi i}{3}}$.
\qed
\end{lemma} 
The above lemma is from  Lemma 7 and Theorem 8 of the paper \cite{Liang2019}. Next, we provide our proof.

\begin{theorem}
    Neither of the two CHMs in \eqref{M23} belongs to any CHM trio.
\end{theorem}
\begin{proof}
    Case 1. The matrix \begin{eqnarray}M_1=\begin{bmatrix}
i & 1 & 1 & 1 & 1 & 1\\
1 & i & 1 & 1 & -1 & -1\\
1 & 1 & i & -1 & 1 & -1\\
1 & 1 & -1 & i & -1 & 1\\
1 & -1 & 1 & -1 & i & 1\\
1 & -1 &-1 & 1 & 1 & i
\end{bmatrix}\end{eqnarray} is complex equivalent to 
\begin{eqnarray}
    D_0=\begin{bmatrix}
        1 & 1 & 1 & 1 & 1 & 1  \\
        1 & -1& i &-i &-i & i  \\
        1 & i &-1 & i &-i &-i  \\
        1 &-i & i &-1 & i &-i  \\
        1 &-i &-i & i &-1 & i  \\
        1 & i &-i &-i & i &-1  
\end{bmatrix}.
\end{eqnarray}
The part $\mathrm{IV.A.}Special\  Hadamard\  matrices$ of paper \cite{bw09} shows that any CHM trio does not contain the matrix $D_0$. Hence any CHM trio does not contain the matrix $M_1$ under complex equivalence.\par
  Case 2. The matrix \begin{eqnarray}M_2=\begin{bmatrix}
\omega & \omega & 1 & 1 & 1 & 1\\
\omega & -\omega & -1 & 1 & -1 & 1\\
1 & 1 & \omega & \omega & 1 & 1\\
1 & -1 & -\omega & \omega & -1 & 1\\
1 & 1 & 1 & 1 & \omega &\omega\\
-1 & 1 &1 & -1 & \omega & -\omega
\end{bmatrix}\end{eqnarray}
contains a submatrix 
$\begin{bmatrix}
    \omega & 1 & 1\\
    1 & \omega & 1\\ 
    1 & 1 & \omega
\end{bmatrix}$ which is a $3\times 3$ CHM. By Lemma 11(v) in \cite{Chen2017Product},  any CHM trio does not contain the matrix $M_2$ under complex equivalence.
\end{proof}


\section{Theorem 2 and Theorem 3 in \cite{commentMUB}}

For the Theorem 2 and Theorem 3 mentioned in \cite{commentMUB}, we show that they are equivalent. By the claims of Lemma 1 and Theorem 2, one can see that the latter implies the former. 
So it suffices to prove Theorem 2 or 3. In order to remedy them as much as possible, we provide new results as follows.
\begin{lemma} 
\label{le:17}
If $H$ is an $H_2$-reducible matrix and the number of  $2\times 2$ complex Hadamard submatrix of $H$ is $N>9$, then $N\geq 17$.
\end{lemma}
\begin{proof}
    Up to complex equivalence, we can assume that
    \begin{eqnarray}
        H=[h_{ij}]=\begin{bmatrix}
            A_1 & A_2 & A_3 \\
            A_4 & A_5 & A_6 \\
            A_7 & A_8 & A_9 
        \end{bmatrix}=\begin{bmatrix}
1 & 1 & 1 & 1 & 1 & 1\\
1 & -1 & h_{23} & -h_{23} & h_{25} & -h_{25}\\
1 & h_{32} & h_{33} & h_{34} & h_{35} & h_{36}\\
1 & -h_{32} & h_{43} & h_{44} & h_{45} & h_{46}\\
1 & h_{35} & h_{53} & h_{54} & h_{55} & h_{56}\\
1 & -h_{35} & h_{63} & h_{64} & h_{65} & h_{66}\\
\end{bmatrix},
    \end{eqnarray}
    and $A_k$($k$=1,...,9) are $2\times 2$ CHMs. Suppose $N>9$, without loss of generality one can assume that $h_{mn}=-1,m>2$, then the submatrix $\begin{bmatrix}
        1 & 1 & 1 & 1 & 1 & 1\\
        1 & h_{m2} & h_{m3} & h_{m4} & h_{m5} & h_{m6}\\
    \end{bmatrix}$ has at least three $2\times 2$ complex Hadamard submatrices. On the other hand, there is a complex Hadamard submatrix $S_2=\begin{bmatrix}
        1 &  h_{sn}\\
        1 &  h_{tn}
    \end{bmatrix}$ such that $\{s,t\}\neq \{1,2 \},\{3,4 \},\{5,6 \}$, the submatrix $\begin{bmatrix}
        1 & h_{s2} & h_{s3} & h_{s4} & h_{s5} & h_{s6}\\
        1 & h_{t2} & h_{t3} & h_{t4} & h_{t5} & h_{t6}\\
    \end{bmatrix}$ has at least three $2\times 2$ complex Hadamard submatrices. Hence $N\geq 9+3+3=15$.\par
    Further, if $n>2$, let $\{u,v\}=\{1,2,...,6\}-\{1,m,s,t\}$, then the submatrix $\begin{bmatrix}
        1 & h_{u2} & h_{u3} & h_{u4} & h_{u5} & h_{u6}\\
        1 & h_{v2} & h_{v3} & h_{v4} & h_{v5} & h_{v6}\\
    \end{bmatrix}$ has at least five $2\times 2$ complex Hadamard submatrices. Hence $N\geq 9+3+3+5-3=17$. Otherwise $n=2$, $H$ has a $3\times 2$ real submatrix with rank two under complex equivalence. Thus $H$ has the form of \eqref{23} under complex equivalence, and has at least seventeen $2\times 2$ complex Hadamard submatrices. So we complete this proof.
\end{proof}

One can verify that the following matrix has at least seventeen $2\times 2$ sub-CHMs by the software Mathematica.  When the parameters $z_1,z_2$ are suitable (for example, $z_1=e^i,z_2=e^{0.5i}$), the number seventeen can be reached. 
\begin{eqnarray}
\label{23}
H=[h_{ij}]=\begin{bmatrix}
1&1&{1}&{1}&{1}&{1}\\
1& -1 & {z_1} & {-z_1} & {z_1} & {-z_1} \\
{1} &{z_2} & -f_1 & -z_2f_2 & -{f_3}^* & -z_2{f_4}^* \\
{1} &{-z_2}& -z_1{f_2}^* & z_1z_2{f_1}^* & -z_1f_4 & z_1z_2f_3 \\
{1}&{z_2}&-{f_3}^*&-z_2{f_4}^*&-f_1&-z_2f_2\\
{1}& {-z_2} & -z_1f_4 &z_1z_2f_3 &-z_1{f_2}^* & z_1z_2{f_1}^*
\end{bmatrix},
\end{eqnarray}
where $z_1=e^{\mathrm{i}x_1}$, $z_2=e^{\mathrm{i}x_2}$, $-\frac{\pi}{2}<x_1\leq \frac{\pi}{2}$, and $-\frac{\pi}{2}<x_2\leq \frac{\pi}{2}$. Further the functions $f_1,f_2,f_3,f_4$ are given by
\begin{eqnarray}
&f_1=f(x_1,x_2),
\quad f_2=f(x_1,-x_2),&\\
&f_3=f(-x_1,-x_2),\quad f_4=f(-x_1,x_2),&
\end{eqnarray}
where 
\begin{eqnarray}
&& f(x_1,x_2)
\\ 
&=&(1-\frac{1}{2}(1-z_1)(1-z_2))(\frac{1}{2}+i\sqrt{\frac{1}{1-\frac{1}{4}(z_1-{z_1}^*)(z_2-{z_2}^*)}-\frac{1}{4}})  \\
&=& e^{i(x_1+x_2)/2}(\cos (\frac{x_1-x_2}{2})-i\sin (\frac{x_1+x_2}{2}))(\frac{1}{2}+i\sqrt{\frac{1}{1+\sin {x_1}\sin {x_2}}-\frac{1}{4}}).
\notag\\ 
\end{eqnarray}
\begin{lemma}
\label{le:23=17}
    The family in \eqref{23} is also the general expression  of the $6\times 6$ CHMs which have $3\times 2$ real submatrices with rank two under complex equivalence.
\end{lemma}

\begin{proof}
    Suppose $H$ is a $6\times 6$ CHM which has $3\times 2$ real submatrices with rank two. Then  $H$ is complex equivalent to
    \begin{eqnarray}
        \begin{bmatrix}
1&1&{1}&{1}&{1}&{1}\\
1& -1 & {z_1} & {-z_1} & {z_3} & {-z_3} \\
{1} &{z_2} & * & * & * & *  \\
{1} &{-z_2} & * & * & * & * \\
{1}&{z_2} & * & * & * & * \\
{1}& {-z_2} & * & * & * & *
\end{bmatrix},
    \end{eqnarray}
where $*$ represents a complex number of modulus one. Using the properties of $H_2$-reducible matrices one can show that $z_1^2-z_3^2=\mathcal{M}(z_2^2)-\mathcal{M}(z_2^2)=0$, then one can assume that $z_1=z_3$. By continuing to use the properties \cite{karlsson11} of $H_2$-reducible matrices, we can show that $H$ is complex equivalent to the matrix in \eqref{23}. So we have completed this proof.
\end{proof}
If $H$ has a $3\times 2$ real submatrix with rank one, then $H$ is from the Fourier family $F_6^{(2)}$ which have been excluded from the CHM trio. If $H$ has a $3\times 2$ real submatrix with rank two, then Lemma \ref{le:23=17} implies that the number of  $2\times 2$ sub-CHMs of $H$ is more than nine. Hence,  if Theorem 3 in \cite{commentMUB} holds then Lemma 1 in \cite{commentMUB} holds.

\begin{lemma}   If $H$ is an $H_2$-reducible matrix and the number of  $2\times 2$ complex Hadamard submatrices of $H$ is $N$, then $N=9,17$ or $N>18$.
\end{lemma}
\begin{proof}
    First of all, the matrix with parameters $z_1=1, \theta=\phi=3$ from Karlsson family $K_6^{(3)}$ \cite{karlsson11} satisfies $N=9$. The matrix with parameters $z_1=e^i,z_2=e^{0.5i}$ in \eqref{23} satisfies $N=17$. On the other hand, Lemma \ref{le:17} shows that $N\neq 10,...,16$. Next, we assume that the matrix $H$ satisfies $N\geq 18$ and we show that $N\neq 18$.
    \par
    1), if $H$ has a $3\times 2$ real submatrix with rank two under complex equivalence, then $H$ has the form of \eqref{23} under complex equivalence, and has more than seventeen $2\times 2$ complex Hadamard submatrices. Using the properties of $H_2$-reducible matrices and the similar method in Lemma \ref{le:17} one can show that $N\neq 18$.\par
    
2), if $H$ does not have a $3\times 2$ real submatrix with rank two under complex equivalence, then $H$ has the submatrix $\begin{bmatrix}
        1 & 1 & 1\\
        1 & -1&* \\
        1 & * &-1
    \end{bmatrix}$ under complex equivalence. Using Lemma 7 of \cite{xyc} one can show that $H$ has the submatrix $\begin{bmatrix}
        1 & 1 & 1 & 1\\
        1 & -1&*  & *\\
        1 & * &-1 & *\\
        1 & * & * &-1
    \end{bmatrix}$ under complex equivalence. One can verify that $N\neq 18$ by the software Mathematica.\par
    In conclusion, we complete this proof.
\end{proof}

\par
To sum up, we have shown that there does not exist a $6\times 6$ CHM with the number of $2\times 2$ sub-CHMs being $10,...,16,18$. It means that Theorem 2 and 3 in \cite{commentMUB} are equivalent. To prove the 
Theorem 3, one need to prove that the $6\times 6$ CHMs whose number of $2\times 2$ sub-CHMs are $17,19,20,...$ and so on do not belong to any CHM trio. One of such CHMs is the known Hermitian CHMs, which have at least twenty-seven $2\times 2$ sub-CHMs (one can verify this by the software Mathematica).

\section{Lemma 1 in \cite{commentMUB}}

The proof of this lemma has some mistakes. It is difficult to fix them relying solely on the existing techniques. One need to consider some new directions, such as the eigenvalue analysis of matrices, simplification of special polynomial equations, and so on.

\section{Conclusion}

We have partially addressed the concerns raised by Daniel McNulty and Stefan Weigert regarding our previous work. For Theorem 1 in \cite{commentMUB}, we have provided a new proof to confirm its correctness. Regarding Theorem 2 and Theorem 3 in \cite{commentMUB}, we have shown that they are equivalent supported by new results, say the number
of $2\times 2$ CHMs of any $H_2$-reducible matrix is not 10, ..., 16, 18. Further, Theorem 2 implies Lemma 1 in \cite{commentMUB}. Unfortunately we are not able to save Theorem 2 and Lemma 1 at present. Nevertheless, we believe that our work (including other correct results not mentioned in \cite{commentMUB}) is still of practical usefulness for characterizing theoretical quantum-information problems such as CHMs and MUBs, and so on.

\section*{Acknowledgements}
We would like to thank the comments of Daniel McNulty and Stefan Weigert. The main authors of this paper are Mengfan Liang and Lin Chen, professor Li Yu did not participate. We finished this work on the basis of 'Product states and Schmidt rank of mutually unbiased bases in dimension six' whose authors are Lin Chen and Li Yu.
Mengfan Liang and Lin Chen were supported by the NNSF of China (Grant No.12471427), and the Fundamental Research Funds for the Central Universities (Grant No. 4303088).

\bibliographystyle{unsrt}

\bibliography{mengfan}

\end{document}